\documentstyle[12pt
]{article}

\renewcommand{\baselinestretch}{1.5}

\def\beq{\begin{equation}}
\def\eeq{\end{equation}}
\def\bea{\begin{eqnarray}}
\def\nn{\nonumber \\ }
\def\eea{\end{eqnarray}}
\def\ds{\displaystyle}

\def\ms{\medskip}
\def\bs{\bigskip}
\def\ni{\noindent}

\def\req#1{(\ref{#1})}
\def\scaption#1{\caption{\protect\small \protect\baselineskip1.5ex #1}}
\def\ie{{\it i.e.}\ }
\def\eg{{\it e.g.}\ }
\def\ctg{{\rm ctg}\ }

\begin{document}

\hfill UB-ECM-PF 92/33

\hfill December 1992

\vspace*{3mm}

\begin{center}

{\LARGE \bf
Spectral Zeta Functions for Spherical Aharonov-Bohm Quantum Bags}

\vspace{4mm}

\renewcommand
\baselinestretch{0.8}
{\sc E. Elizalde}\footnote{E-mail address: eli @ ebubecm1.bitnet} and
{\sc S. Leseduarte}
\\
{\it Departament E.C.M., Facultat de F!sica, Universitat de
Barcelona, \\
Diagonal 647, 08028 Barcelona, Spain} \\  and \\
{\sc A. Romeo}
 \\ {\it Departament de Matemtica Aplicada i Anlisi, \\
Facultat de Matemtiques, Universitat de Barcelona, \\
 Gran Via de les Corts Catalanes 585, 08071 Barcelona, Spain}
\ms

\renewcommand
\baselinestretch{1.4}

\vspace{5mm}

{\bf Abstract}

\end{center}

We  study the sum $\ds\zeta_H(s)=\sum_j E_j^{-s}$
over the
eigenvalues $E_j$ of the Schrdinger equation in a  spherical
domain with Dirichlet walls, threaded by a line of magnetic flux.
Rather than using Green's function techniques,
we tackle the mathematically nontrivial problem of finding
exact sum rules for the zeros of Bessel functions $J_{\nu}$,
which are extremely helpful when seeking numerical approximations to
ground state energies.
These results are particularly valuable
if $\nu$ is neither an integer nor half an odd one.
\vspace{20mm}

\hfill  Dedicated to David Bohm, {\it in memoriam}

\section{Introduction}

There is a large class of physically interesting problems where the
Hamiltonian is closely related to the operator of Bessel's
equation, this includes, \eg: \ (i)
 classical equations for vibrating strings and drumheads, heat
conduction in cylinders, normal modes in resonant cavities,
Fraunhofer difraction through circular apertures, etc.; \ (ii)
 quantum particles which move freely within cylindrically or spherically
symmetric domains, on whose boundaries the wave function vanishes
(they play the role of reflecting or bouncing walls).
The two-dimensional case has been made further complicated in two
different ways:
by altering the shape of the boundary ---quantum billiards
\cite{It,Be}--- and by introducing magnetic potentials
such that the domain is threaded by a line of magnetic flux
---Aharonov-Bohm quantum billiards \cite{Be}.

This paper deals with the extension of the last situation to spherical
domains of arbitrary radius, $a$, diametrically threaded by a flux
line.
For such a system, the time-independent Schrdinger equation
takes the form
\beq {1 \over 2 \mu} [-i\hbar {\bf\nabla} -q {\bf A}({\bf r}) ]^2
\Psi({\bf r})= E \Psi({\bf r}), \label{Seq}\eeq
where $\mu$ is the mass of the particle, $q$ its charge and
${\bf A}({\bf r})$ is some suitable vector potential, which produces a
straight line of flux along the $z$-axis. To this end we choose
\beq {\bf A}({\bf r})={ \Phi \over 2\pi } {1 \over r \sin \theta }
{\bf e}_{\varphi}, \eeq
where $\Phi$ is the value of the flux and ${\bf e}_{\varphi}$
denotes
the unitary tangent vector for increasing $\varphi$ (the usual spherical
coordinates, $r, \theta$ and $\varphi$, are employed).
This corresponds to selecting a gauge in which the lines of the vector
potential become concentric circles.
After noticing that
${\bf \nabla} \cdot {\bf A}=0$, as should be expected, \req{Seq} reduces
to
\beq \left[ \nabla^2 +k^2 -{1 \over r^2 \sin^2 \theta}
\left( 2 i \alpha {\partial \over \partial \varphi} +\alpha^2 \right)
\right] \Psi( r, \theta, \varphi )=0, \eeq
where $\ds k^2= {2 \mu E \over \hbar^2}$ and
$\ds\alpha={q \over \hbar}{\Phi \over 2 \pi}$. Applying  the usual
variable separation method, we set
$\Psi( r, \theta, \varphi )=R(r) {\cal Y}( \theta, \varphi )$ and arrive
at
\bea\ds \left[
{\partial \over \partial r} r^2 {\partial \over \partial r}
+k^2 r^2 - C \right] R(r)&=&0, \label{radeq} \\
\ds\left[ {1 \over \sin\theta } {\partial \over \partial\theta }
\sin\theta {\partial \over \partial\theta }
+{1 \over \sin^2\theta } \left( {\partial^2 \over \partial \varphi^2}
-2i\alpha {\partial \over \partial \varphi} -\alpha^2 \right) +C
\right] {\cal Y}( \theta, \varphi )&=&0, \label{angeq}\eea
with $C$ a separation constant. Writting $C=\beta(\beta
+1)$,
\req{radeq} becomes the spherical Bessel equation of parameter $\beta$,
not necessarily integer. Thus, the solution regular at $r=0$ is
\beq R(r) \sim j_{\beta}(kr)= \sqrt{ \pi \over 2kr } J_{\beta+1/2}(kr),
\eeq
where $j_{\beta}$ denotes the spherical Bessel function.
As a result of the boundary condition imposed on the surface $r=a$,
$ka$ must be a zero of $J_{\beta+1/2}$. The energy is therefore
discretized in the way
\beq E_{\beta n}={ \hbar^2 \over 2 \mu a^2 } j_{\beta+1/2, n}^2, \eeq
$j_{\beta+1/2, n}$ being the $n$th positive zero of $J_{\beta+1/2}$.

Taking ${\cal Y}(\theta, \varphi)=\Theta(\theta)\phi(\varphi)$ with
the function $\phi(\varphi)$ of the form $e^{im\varphi}$, $m \in {\bf
Z}$, and doing the change of variables $x=\cos\theta$, eq. (\ref{angeq})
yields
\beq \left[ (1-x^2) {d^2 \over dx^2} -2x {d \over dx} + \beta( \beta+1 )
-{ (m-\alpha)^2 \over 1-x^2 } \right] \Theta(\theta(x))=0, \eeq
and writting $\Theta(\theta (x)) = (1-x^2)^{|m-\alpha|/2} u(x)$, the
function $u(x)$ satisfies the following equation
\beq
(1-x^2) \ u''-2x \, (|m-\alpha|+1) \ u'+ \left[ C-|m-\alpha|
(|m-\alpha|+1)\right] \, u=0. \eeq
 Since $x=0$ is a
regular point, we make
the assay $\ds u(x)= \sum_{k=0}^{\infty} a_k x^k$, which leads to the
recurrence \beq a_{k+2}=
{ (k+|m-\alpha|) (k+|m-\alpha|+1) -C \over (k+2)(k+1) }
a_k . \eeq
Good behaviour at $x=\pm1$ can only be achieved by truncating the series
into a polynomial. This takes place if there is some  $p\in $ {\bf N}
such that $C=(|m-\alpha|+p) (|m-\alpha|+p+1)$. This recurrence would be
the same as the one for
Legendre's associated functions $P_l^m$, were it not for the presence
of a (generally) {\it noninteger} $\alpha$. Therefore, we adopt the
notation
\beq \Theta(\theta) \sim P_{\beta}^{m-\alpha}( \cos \theta ),
\ |m-\alpha| \le \beta, \eeq
where $\beta \equiv |m-\alpha|+p$ is also non-integer, as a rule.
The general form of the wave function is a superposition of the sort
\beq \Psi(r, \theta, \varphi )=\sum_{n,l,m} c_{nlm} \
j_{ \beta }\left( j_{ \beta+1/2, n } {r \over a}
\right)
P_{ \beta }^{m-\alpha}( \cos \theta ) \ e^{im \varphi } . \eeq
The angular eigenfunctions
$\sim P_{\beta}^{m-\alpha}( \cos \theta ) \ e^{im \varphi }$
generalize the spherical harmonics $Y_{lm}(\theta, \varphi)$,
corresponding to $\alpha=0$. It is precisely for this reason
---\ie the presence of $\alpha$--- that the solutions are no longer
eigenfunctions of the angular momentum ${\bf L}^2$ (they are instead
eigenfuctions of the mechanical momentum $\Pi^2$).
This was to be expected, as radial symmetry has been broken by ${\bf
A}({\bf r})$.
The energy spectrum has now turned into
\beq E_{\beta,n}={\hbar^2 \over 2 \mu \alpha^2}
j_{ \beta+1/2, n }^2, \ m \in {\bf Z}, \ \beta =|m-\alpha|+p, \ p \in
{\bf N}. \eeq
The ground state energy is therefore
$E_{0,1}={\hbar^2 \over 2 \mu \alpha^2}j_{ | \alpha |+1/2, 1 }^2$, for
$|\alpha| \leq 1/2$. Notice that $\alpha=1/2$ gives rise to double
degeneracy, since
states with $m$ and $-m+1$ have the same energy. This is also true, in
particular, for the ground state.

Let us recall that the zeta function of a Hermitean operator $A$ with
infinitely many discrete eigenvalues $\lambda_i$ is defined as
\beq \zeta_A(s)=\sum_i {1 \over \lambda_i^s} = {\rm Tr} A^{-s} . \eeq

An alternative to the computer evaluation of these Bessel function zeros
is their approximation by means of sum rules. When $A$ has a discrete
spectrum $ \lambda_1 < \lambda_2 < \lambda_3 < \dots$,
several ways to find estimates of low-lying eigenvalues
---quoted in \cite{Wa}--- are in order. The most adequate here
%
%
is Euler's method (originally applied to the zeros of $J_0$),
which is based on the inequalities
\beq \zeta_A(s)^{-1/s} < \lambda_1 < {\zeta_A(s) \over \zeta_A(s+1)} .
\label{ineq}\eeq
In fact,
$\ds\lim_{s\to\infty} [ \zeta_A(s) ]^{-1/s}
=\lim_{s\to\infty} {\zeta_A(s) \over \zeta_A(s+1)}=\lambda_1$ and, prior
to taking the limit, \req{ineq} always holds.


In our case, we will take $A$ to be the operator of the spherical
Bessel equation \req{radeq} with $C\equiv (|m-\alpha|+p) ( |m-\alpha| +p
+1)$. Thus, for each possible $l$, the zeta function $\zeta_A(s)$
to be considered is of the type $\zeta_A(s) \sim \zeta_{\nu}(2s)$,
with
\beq \zeta_{\nu}(s)=\sum_{n=1}^{\infty} {1 \over j_{\nu n}^s },
\ {\rm Re}\ s >1, \label{defzetanu}\eeq
where $j_{\nu n}$ is the $n$th nonvanishing zero of $J_{\nu}$.
The condition $\ {\rm Re}\ s >1$ stems from considering the asymptotic
form of the $j_{\nu n}$'s, which is roughly
$\left( \nu-{1 \over 2} \right){\pi \over 2}+\pi n$, and from comparing
with
the Hurwitz zeta function. At any rate, the existence of poles at some
values of $s$ ought to be checked by operator heat-kernel expansion
methods.
\bs

\section{Recursive Rules for Zeta Functions associated to Bessel
Operators}

\subsection{Quadratic Recursion}

 When written in the
form of an infinite product involving its nonvanishing zeros,
the $\nu$th Bessel function of the first species
reads
\beq J_{\nu}(z)= { z^{\nu} \over 2^{\nu} \Gamma( \nu +1) }
\prod_{n=1}^{\infty} \left( 1-{z^2 \over j_{\nu n}^2} \right) .
\eeq
Differentiating the $\log$ of this expression with respect to $z$, and
using the identity
\beq J_{\nu}'(z)={\nu \over z}J_{\nu}(z)-J_{\nu+1}(z), \label{BFIi}\eeq
we arrive at
\beq {J_{\nu +1}(z) \over J_{\nu}(z)}=
2z\sum_{n=1}^{\infty} {1 \over j_{\nu n}^2 -z^2}  \eeq
and, since $j_{\nu 1} < j_{\nu 2} <...$, taking
$|z| < j_{\nu 1}$ we are able to put
$\ds{1 \over 1-\left( z \over j_{\nu n} \right)^2}
=\sum_{m=0}^{\infty} \left( z \over j_{\nu n} \right)^{2m}.$
Then, interchanging the summations and looking at \req{defzetanu}
we come to
\beq {J_{\nu +1}(z) \over J_{\nu}(z)}=
2 \sum_{n=1}^{\infty} \zeta_{\nu}(2n) z^{2n-1} . \label{start} \eeq
This formula will constitute our starting point and is a well
 known result (see \eg \cite{Bate}, pg. 61).
Now one can, of course, find the values of the zeta functions
through
$\ds \zeta_{\nu}(2n)
={1 \over 2 (2n-1)!} \left. {d^{2n-1} \over dz^{2n-1}}
{J_{\nu +1}(z) \over J_{\nu}(z)}\right\vert_{z=0}$.
Rather than doing this, we take the derivative of \req{start} with
respect to $z$, and making use of \req{BFIi} and of
$\ds J_{\nu + 1}'(z)= J_{\nu}(z)-{\nu +1 \over z} J_{\nu+1}(z)$, we
obtain
\beq 1-{2\nu+1 \over z}{J_{\nu +1}(z) \over J_{\nu}(z)}
+\left(J_{\nu +1}(z) \over J_{\nu}(z)\right)^2
=\sum_{n=1}^{\infty} \zeta_{\nu}(2n) \ 2(2n-1) z^{2n-2} . \eeq
Replacing ${J_{\nu +1}(z) \over J_{\nu}(z)}$ with the r.h.s. of
\req{start}, we get a sum of series which can be recast into the form
\beq 1+\sum_{n=1}^{\infty}\left[ -4(\nu+n) \zeta_{\nu}(2n)
+4\sum_{l=1}^{n-1} \zeta_{\nu}(2l) \zeta_{\nu}(2n-2l) \right] z^{2n-2}
=0. \eeq
Since it must be satisfied for any $z$ with $|z|< j_{\nu 1}$, the
identity must hold separately for every power of this variable.
In particular, for $n=1$ we are led to
\beq \zeta_{\nu}(2)={1 \over 4(\nu +1)}, \label{zetanu2} \eeq
while, for $n \ge 2$, the `quadratic' recursive relation
\beq \zeta_{\nu}(2n)=
{1 \over n+\nu} \sum_{l=1}^{n-1} \zeta_{\nu}(2l) \zeta_{\nu}(2n-2l)
\label{rec2} \eeq
follows. In Appendix A we present an alternative derivation of this
formula by direct integration over a convenient contour in the
complex plane.

This equation, together with the previous, provides the value
of $\zeta_{\nu}(2n)$ at any even integer argument. The first results
thus obtained are:
\bea
\zeta_{\nu}(4)&=&\ds{1 \over 2^4 (\nu +1)^2 (\nu +2)},
\label{zetanu4} \\
\zeta_{\nu}(6)&=&\ds{1 \over 2^5 (\nu +1)^3 (\nu +2) (\nu +3)},
\label{zetanu6} \\
\zeta_{\nu}(8)&=&\ds{5 \nu +11
\over 2^8 (\nu +1)^4 (\nu +2)^2 (\nu +3) (\nu +4)},
\label{zetanu8} \\
\zeta_{\nu}(10)&=&\ds{7 \nu +19
\over 2^9 (\nu +1)^5 (\nu +2)^2 (\nu +3) (\nu +4) (\nu +5)},
\label{zetanu10} \\
&\vdots& \nonumber
\eea
These expressions are quoted in \cite{Wa} as due to Rayleigh.
Now, with the help of \req{rec2}, their derivation has been
straightforward.

An immediate consequence of \req{rec2} is the
nontrivial equality
\beq \sum_{l=1}^{n-1} \sum_{k,m=1 \atop k \ne m}^{\infty}
{1 \over j_{\nu k}^{2l} j_{\nu m}^{2n-2l} }=
(\nu+1)\zeta_{\nu}(2n) . \eeq

The procedure here described finds  application to other functions
that can be expressed as infinite products. For instance, we take
$\ds \sin z=
z \prod_{n=1}^{\infty}\left( 1-{z^2 \over \pi^2 n^2}\right)$.
Its logarithmic derivative gives (assuming $|z| < \pi$)
a known expansion for $\ctg z$:
\beq \ctg z={1 \over z}
-2\sum_{m=0}^{\infty} \zeta(2m+2) {z^{2m+1} \over \pi^{2m+2} },
\label{expctg} \eeq
where $\zeta$ is the Riemann zeta function itself.
Differentiating with respect to $z$, the l.h.s. will be
$-\csc^2 z=-(1+\ctg^2 z)$. Replacing $\ctg z$ with the series
\req{expctg}, and the r.h.s. with its derivative, we obtain an
identity which must be fulfilled for every power of $z$.
Thus
\beq
\left\{\begin{array}{lll} \zeta(2)&=&\ds{\pi^2 \over 6}, \nn
\zeta(2n)&=&\ds{1 \over n+{1 \over 2}}
\sum_{l=1}^{n-1} \zeta(2n-2l) \zeta(2l), \ n \ge 2. \end{array}\right.
\label{rec2sin}\eeq
This provides a method for the recurrent evaluation of $\zeta(2n)$
from just the knowledge of $\zeta(2)$.
Bearing in mind the relation between zeta function and Bernoulli
numbers,
$\ds \zeta(2n)=(-1)^{n+1}{ (2\pi)^{2n} \over 2 (2n)! } B_{2n}$,
the previous equations read
\beq
\left\{ \begin{array}{lll} B_2&=&\ds{1 \over 6}, \nn
B_{2n}&=&\ds-{1 \over 2n+1} \sum_{l=1}^{n-1} \pmatrix{ 2n \cr 2l }
B_{2n-2l} B_{2l} , \ n\ge 2 . \end{array}\right.
\eeq
They supply a different way of finding all the nonvanishing Bernoulli
numbers ($B_1$ aside) from the value of $B_2$.
Eq. \req{rec2sin} yields also
\beq \sum_{l=1}^{n-1} \sum_{k,m=1 \atop k \ne m}^{\infty}
{1 \over k^{2l}} {1 \over m^{2n-2l}}=
{3 \over 2} \zeta(2n) . \eeq
Actually, all these results can as well be obtained from those for
$\zeta_{\nu}$ by taking $\nu={1 \over 2}$, since
$J_{1/2}(z)=\sqrt{2 \over \pi z} \sin z$.
\ms




\subsection{Linear Recursion}

The above recurrence has turned out to be a powerful tool for
successively obtaining the zeta functions of even argument. But, having
succeeded in finding a `second order' law ---as the r.h.s involves
products of zeta functions at lower arguments--- we would also like to
have a linear rule available. The example of Euler's method (shown, \eg,
in \cite{Wa}, pg. 500) takes advantage of a {\it linear} recurrence
among
the $\zeta_{0}(2n)$'s. Now, our aim is   to find the general form of
such type of relation, thus extending the procedure to any $\nu$.

We start, once more, by taking \req{start}, this time written as
\beq J_{\nu+1}(z) =
2J_{\nu}(z) \sum_{n=0}^{\infty} \zeta_{\nu}(2n+2)z^{2n+1} . \eeq
Next, replacing the $J_{\nu}$'s with their power series expansions
\beq J_{\nu}(z)= \sum_{k=0}^{\infty}{ (-1)^k \over k! (k+\nu )! }
\left( z \over 2 \right)^{2k+\nu} \eeq
(in general, these factorials are to be understood as gamma functions $(
k+\nu )! \equiv \Gamma( k+\nu +1 )$), we obtain a relationship
between series which, after some index rearrangements, reads
\beq \sum_{n=0}^{\infty} { (-1)^n \over 2^{2n+\nu} }
\left[ {1 \over 2 \ n! (n+\nu+1)!} -
2 \sum_{k=0}^n { (-1)^k 2^{2k} \over (n-k)! (n-k+\nu)! }
\zeta_{\nu}(2k+2) \right] z^{2n+\nu+1}= 0. \eeq
Validity for any $z$ calls for the vanishing of every coefficient.
Therefore
\beq {1 \over 4(n+\nu+1)}=\sum_{k=0}^n (-1)^k 4^k
{ n! \over (n-k)! } { (n+\nu)! \over (n+\nu-k)! } \zeta_{\nu}(2k+2), \
n \ge 0, \eeq
which amounts to
\beq {1 \over 4(\nu +n)}
=\ds\sum_{k=0}^{n-1} (-1)^k 4^k (k!)^2
\pmatrix{ n-1 \cr k } \pmatrix{ \nu+n-1 \cr k } \zeta_{\nu}(2k+2), \
n\ge 1. \label{lrec}\eeq
This is the linear relation we were looking for.

The first resulting identities are:
\bea {1 \over 4(\nu +1)}&=&\zeta_{\nu}(2), \\
{1 \over 4 (\nu + 2)}&=&\zeta_{\nu}(2)-4(\nu +1)\zeta_{\nu}(4), \\
{1 \over 4 (\nu +3) }&=&
\zeta_{\nu}(2) - 8(\nu +2) \zeta_{\nu}(4)
+32(\nu+1)(\nu+2) \zeta_{\nu}(6), \\
{1 \over 4 (\nu +4) }&=&
\zeta_{\nu}(2) -12(\nu +3) \zeta_{\nu}(4)
+96(\nu+2)(\nu+3) \zeta_{\nu}(6) \nn
&&-384(\nu+3)(\nu+2)(\nu+1) \zeta_{\nu}(8), \\
{1 \over 4 (\nu +5) }&=&\ds
\zeta_{\nu}(2) -16(\nu +4) \zeta_{\nu}(4)
+192(\nu+4)(\nu+3) \zeta_{\nu}(6) \nn
&&\ds-1536(\nu+4)(\nu+3)(\nu+2) \zeta_{\nu}(8)
+6144(\nu+4)(\nu+3)(\nu+2)(\nu+1) \zeta_{\nu}(10). \nn
\label{lrec10} \eea
This can be regarded as a system of linear equations in
$\zeta_{\nu}(2), \zeta_{\nu}(4), \dots, \zeta_{\nu}(10)$,
which can be solved
by repeated substitution from the first. It can be checked that its
solution is \req{zetanu2}, \req{zetanu4}, \req{zetanu6}, \req{zetanu8},
\req{zetanu10}. The following remark is in order: although a linear
recursion looks in principle simpler than one involving products,
in practice it is much easier to work with \req{rec2} than with
\req{lrec}.
 \bs

\section{Numerical evaluation of the ground state energies}

Euler's method, combined with \req{zetanu2},\req{rec2}, turns out to
be a very efficient procedure for the computation of $j_{\nu,1}$,
associated to the ground state (here $\nu=|m-\alpha|+p+1/2$).
The lower and upper bounds in \req{ineq} are now
$[ \zeta_{\nu}(2k) ]^{-1/(2k)}$
and $\sqrt{\zeta_{\nu}(2k)/\zeta_{\nu}(2k+2)}$, respectively,
where the square root
comes from the factor 2 in the argument when going from $\zeta_{\nu}$
to $\zeta_A$.
We calculate them for increasing $k$'s until convergence of both the
lower and the upper successions becomes apparent.
At every step, a new $\zeta_{\nu}(2k)$
is found by means of a recursive function ---implementing
\req{zetanu2} and \req{rec2}--- which is part of a simple C program.
Table \ref{onlytable} shows the intermediate results
obtained for the cases $\nu=0,1/2$ and $1$, whose first nonvanishing
zeros are known to be $2.404826\dots$, $\pi$ and $3.831706\dots$,
respectively.
The figures obtained after
arriving at $k=10$ for different $\nu$'s between 0 and 1 are
displayed in the right box.
Of these results, only those for
$\nu \ge 1/2$ can be physically meaningful, as the lowest possible
ground state corresponds to $\nu=1/2$. Such value would be
attained only for $\alpha=0$, while higher $\nu$'s correspond
to nonvanishing $\alpha$'s.

The calculation has been repeated ---with the same accuracy---
using a program which implements the linear recursion instead of the
quadratic one. No figure of the results shown in Table 1 has
changed, alhough the execution time is now shorter.
However, away from this range the
numerical errors produced by both algorithms may differ. Actually,
for $\nu$ close to 100, we have observed differences of the orders of
$10^{-6}$ and even $10^{-5}$ within the first ten steps.

\begin{table}[htbp]
\begin{center}
\begin{tabular}{cc}
\begin{tabular}{|r|cc|cc|cc|}
\hline\hline
$k$&\multicolumn{2}{c|}{$\nu=0$}
   &\multicolumn{2}{c|}{$\nu=1/2$}
   &\multicolumn{2}{c|}{$\nu=1$} \\ \hline
1  & 2.000000&2.828427 & 2.449490&3.872983 & 2.828427&4.898979 \\
2  & 2.378414&2.449490 & 3.080070&3.240370 & 3.722419&4.000000 \\
3  & 2.401874&2.412091 & 3.132603&3.162278 & 3.812737&3.872983 \\
4  & 2.404424&2.406133 & 3.139995&3.146427 & 3.827710&3.843076 \\
5  & 2.404766&2.405069 & 3.141280&3.142768 & 3.830778&3.834980 \\
6  & 2.404816&2.404871 & 3.141528&3.141883 & 3.831478&3.832667 \\
7  & 2.404824&2.404834 & 3.141579&3.141665 & 3.831648&3.831990 \\
8  & 2.404825&2.404827 & 3.141590&3.141611 & 3.831691&3.831791 \\
9  & 2.404826&2.404826 & 3.141592&3.141597 & 3.831702&3.831731 \\
10 & 2.404826&2.404826 & 3.141593&3.141594 & 3.831705&3.831713 \\
\hline\hline \end{tabular}
&
\begin{tabular}{|l|c|}
\hline\hline
$\nu$&$j_{\nu 1}$ \\ \hline
0  &2.404826 \\
0.1&2.557451 \\
0.2&2.707073 \\
0.3&2.854097 \\
0.4&2.998849 \\
0.5&3.141593 \\
0.6&3.282545 \\
0.7&3.421890 \\
0.8&3.559780 \\
0.9&3.696347 \\
1  &3.831705 \\ \hline\hline
\end{tabular}
\end{tabular}
\end{center}
\scaption{ Left: successive values in the approximation to $j_{\nu 1}$,
corresponding to $\nu=0,1/2,1$.
For each $\nu$, the values on the left
and right columns are the lower and upper bounds,
obtained as $\left[ \zeta_{\nu}(2k) \right]^{-1/(2k)}$ and
$\left[ \zeta_{\nu}(2k) / \zeta_{\nu}(2k+2) \right]^{1/2}$
respectively.
Right: different $j_{\nu 1}$'s for some values of $\nu$ between 0 and
1.}
\label{onlytable}
\end{table}

\section{Conclusions}

We have focused in this paper on the derivation of exact and explicit
formulas
for the zeta function corresponding to the zeros of the Bessel function,
$\zeta_{\nu}(2k)$ (for arbitrary $\nu$),
and in their systematic use for obtaining estimates of ground states of
related Hamiltonian operators.
 The accuracy of the procedure here developed has proven to be
excellent. Yet, the main advantage of the method has not been
exploited fully. By this we mean the possibility of finding such
approximations {\it even in cases where the spectrum does not emerge
from standard special functions}, but $\zeta_A$ can nevertheless be
calculated, at least for some special values \cite{Vo}.

Another subject to be explored is the deformation of the spherical
domain. The corresponding case in two dimensions
 was tackled by
means of conformal mappings between the unit disc and other noncircular
domains \cite{Be}. In three dimensions the difficulty would surely
increase because such powerful complex transformations are lacking.
We think that these subjects are worth studying in further detail.

In particular, here, by dealing with a problem
of known spectrum, we have been able to relate the physical information
concerning the energy levels with some remarkable numerical properties
involving Bernoulli numbers and ordinary Riemann zeta functions.
\vspace{5mm}

\ni{\large \bf Acknowledgments}

This work has been partially supported by DGICYT (Spain), research
project PB90-0022.
\bs

\appendix

\section{Appendix:
 Derivation of the quadratic recurrence by complex
integration}

The zeta function
\beq
\zeta_{\nu} (z) \equiv \sum_{m=1}^{\infty} \lambda_m^{-z}, \ \ \ \
\ J_{\nu} (\lambda_m) =0,
\eeq
can be written as the integral
\beq
\zeta_{\nu} (z)= -\frac{1}{2\pi i} \int_{\cal C} dt \,
\frac{J_{\nu}'(t)}{J_{\nu}(t)} \, t^{-z} = \frac{iz}{2\pi}
\int_{\cal C} dt \, \ln \left[ J(t) \right] t^{-z-1},
\label{a1}
\eeq
where $\cal C$ is the sector-like contour of the complex plane
domain
$|\arg (t)| \leq \theta$, with $0< \theta \leq \pi/2$, and
$|t|\geq \epsilon$, with $0<\epsilon < \lambda_1$ and Re $t >1$.
Recalling that, for $|\arg (w)| <\pi - \delta$ ($\delta >0$) and
$|w| \rightarrow \infty$,
\beq
J_{\nu} (w) \sim \sqrt{\frac{2}{\pi w}} \left[ P_{\nu} (w) \cos
\left( w-(\nu + \frac{1}{2}) \frac{\pi}{2} \right) - Q_{\nu} (w)
\sin \left( w-(\nu + \frac{1}{2}) \frac{\pi}{2} \right)  \right],
\eeq
where $P_{\nu}$ and $Q_{\nu}$ are  series of negative powers in $w$
(we assume that $\nu >-1/2$), and particularizing it for the case
$w=\rho \, e^{\pm i\pi/2}$ ($\rho >0$), after substitution in
expression (\ref{a1}), we obtain
\bea
\zeta_{\nu} (z)&=& \frac{iz}{2\pi}  \left\{ \sum_{\pm} \int_{C_{\pm
(\epsilon)}} dt \,  t^{-z-1} \ln \left[ \sqrt{2\pi t} \exp
\left(\pm i
\left( t- \frac{\pi}{2}(\nu + \frac{1}{2}) \right)\right) J(t)
\right]
\right. \nn \\ &+& 2 \left. \left[ \frac{i \epsilon^{-z+1}}{1-z} +
i \left(\nu + \frac{1}{2} \right) \frac{\pi}{2z} \epsilon^{-z} \right]
\right\},
\label{a2}
\eea
where we have taken here $\theta = \pi/2$ and $C_{\pm} (\epsilon)$
are the two halves of the small circular part of the contour, of
radius
$\epsilon$. The two integrals in (\ref{a2}) are analytical
functions of $z$, while the last term is an explicitly meromorphic
function. We thus see that  $\zeta_{\nu}$ has simple pole at $z=1$
with residue $1/\pi$.

Let us now consider the restriction of $\zeta_{\nu}$ to the strip
$-1 <$ Re $z<0$. The limit $\epsilon \rightarrow 0$ yields the
fundamental expression
\beq
\zeta_{\nu} (z)= \frac{z}{\pi} \sin  \left( \frac{\pi}{2} \,
z\right) \int_0^{\infty} d\rho \,  \rho^{-z-1} \ln \left[
\sqrt{2\pi \rho}\,  e^{-\rho} I_{\nu} (\rho)\right], \ \ \ -1<
\mbox{Re} \, z <0.
\label{a3}
\eeq
It can be continued ---in the value of Re $z$--- to the right
(and also to the left) of the real axis. Adding and subtracting in
the integrand a term like $ e^{-\rho} \sum_{k=0}^{2p} \rho^k (a_k
+ b_k \ln \rho)$, with $a_0\equiv \ln [\sqrt{2\pi} /(2^{\nu} \nu
!)]$, $b_0 \equiv \nu + 1/2$, and adjusting the remaining $a$'s and
$b$'s so that the difference with the integrand in (\ref{a3}) be
\beq
\ln \left[ \sqrt{2\pi \rho}\,  e^{-\rho} I_{\nu} (\rho)\right]
- e^{-\rho} \sum_{k=0}^{2p} \rho^k (a_k + b_k \ln \rho) \simeq
{\cal O} \left(\rho^{2p+1} \ln \rho \right)
\label{a4}
\eeq
 when $\rho \rightarrow 0$, it can be proven that the function
$G_{\nu} (\rho) \equiv \ln \left[ 2^{\nu} \nu ! \rho^{-\nu} e^{-
\rho} I_{\nu} (\rho ) \right]$ satisfies
\beq
G'' + {G'}^2 + G' \left( 2+ \frac{2\nu +1}{\rho} \right) +
\frac{2\nu +1}{\rho} =0,
\eeq
which has a solution of the form $G (\rho)= \sum_{k=1}^{\infty} c_k
\rho^k $, $c_1 =-1$, in a neighborhood of $\rho = 0$. In terms of
$G_{\nu} (\rho )$ and $H_{\nu} (\rho )\equiv e^{\rho} G_{\nu} (\rho
) $, the condition (\ref{a4}) is written as
\beq
H_{\nu} (\rho ) -  \sum_{k=1}^{2p} a_k \rho^k + (e^{\rho} -1) \ln
\left( \frac{\sqrt{2\pi}}{2^{\nu} \nu !} \right) + M_{\nu} (\rho )
\ln \rho  \simeq {\cal O} \left(\rho^{2p+1} \ln \rho \right),
\label{a5}
\eeq
where $M_{\nu} (\rho )$  is analytical around $\rho =0$ and
contains all the information on the $b$ coefficients ---which are
rigorously proven to exist but will not interest us here. The $a$'s
are read off from (\ref{a5}), with the result
\beq
a_k = \frac{1}{k!} \ln \left( \frac{\sqrt{2\pi}}{2^{\nu} \nu !}
\right) +\sum_{j=1}^{k} \frac{c_j}{ (k-j)!}.
\eeq
Back to (\ref{a3}), we see that
\bea
\zeta_{\nu} (z)&=& \frac{z}{\pi} \sin  \left( \frac{\pi}{2} \,
z\right) \int_0^{\infty} d\rho \,  \rho^{-z-1} \left\{ \ln \left[
\sqrt{2\pi \rho}\,  e^{-\rho} I_{\nu} (\rho)\right] - e^{-\rho}
\sum_{k=0}^{2p} (a_k + b_k \ln \rho ) \rho^k  \right\} \nn \\
&+& \frac{z}{\pi} \sin  \left( \frac{\pi}{2} \, z\right)
\sum_{k=0}^{2p} \left[ a_k \Gamma (k-z) + b_k \Gamma' (k-z)
\right],
\eea
with the $a$'s and $b$'s determined as above, provides the desired
analytical extension in the strip $-1 < $ Re $z<2p+1$. After some
straigthforward manipulations we see that for $z$ a positive even
integer, we get
\beq
\zeta_{\nu} (2m ) = (-1)^{m+1} m \, c_{2m}.
\eeq
It is also clear that the $b$'s do not contribute to the expression
in the limit and, by working out the precise value of the $c$'s
from (\ref{a5}), we obtain
$\zeta_{\nu} (2)=[4(\nu +1)]^{-1}$ and
the quadratic recurrence
\beq
\zeta_{\nu} (2(p+1)) = \frac{1}{p+\nu +1} \sum_{k=0}^{p-1}
\zeta_{\nu} (2(k+1)) \zeta_{\nu} (2(p-k)),
\eeq
valid for any integer $p$, $p\geq 1$.

\end{document}